\documentclass[%
 reprint,
 superscriptaddress,
 amsmath,amssymb,
 aps,
 prl, 
]{revtex4-2}

\usepackage{graphicx}
\usepackage{physics}
\usepackage{dcolumn}
\usepackage{bm}
\usepackage[colorlinks=true,
            linkcolor=red,
            citecolor=blue,
            urlcolor=blue,
            pdfborder={0 0 0}]{hyperref}

\begin{document}

\title{Sub-second A-scan Acquisition Using Marginal Spectral-Domain Quantum Optical Coherence Tomography}

\author{P. D. Yepiz-Graciano}
\affiliation{Instituto de Ciencias Nucleares, Universidad Nacional Autónoma de México, CDMX 04510, México.}
\email{pablodyg@gmail.com}
\author{D. Salamanca-Roldán}
\affiliation{Centro de Investigaciones en Óptica A.C., León, Guanajuato 37150, México.}
\author{H. Cruz-Ramírez}
\affiliation{Instituto de Ciencias Nucleares, Universidad Nacional Autónoma de México, CDMX 04510, México.}
\author{A. B. U'Ren}
\affiliation{Instituto de Ciencias Nucleares, Universidad Nacional Autónoma de México, CDMX 04510, México.}
\author{R. Ramírez-Alarcón}
\affiliation{Centro de Investigaciones en Óptica A.C., León, Guanajuato 37150, México.}


\begin{abstract} 
We report an optimized implementation of spectral-domain quantum optical coherence tomography (SD-QOCT) capable of acquiring axial scans (A-scans) of multilayer samples in the absence of mechanical scanning, at an unprecedented speed. Building on our group’s earlier work, we demonstrate a proof-of-concept system that integrates a diffraction grating, a high-resolution intensified CCD camera, and a high-flux photon-pair source operating in the VIS–NIR region ($\approx$800 nm). This configuration enables the acquisition of an entire marginal SD-QOCT interferogram in a single camera exposure, yielding a transverse A-scan with a record acquisition time of 100 ms and a penetration depth of $\approx$4 mm. The measured interferometric response shows excellent agreement with the theoretical model. These results represent a decisive step toward the practical deployment of SD-QOCT as a competitive imaging modality for biomedical applications.
\end{abstract}


\maketitle


\section{Introduction}
Quantum optical coherence tomography (QOCT) is the quantum analogue of optical coherence tomography (OCT) \cite{Huang1991}. In QOCT, the classical low-coherence source is replaced by a source of entangled photon pairs;  interference at a beamsplitter gives rise to the Hong–Ou–Mandel (HOM) effect \cite{Hong1987}. Each interface within the sample generates a distinct HOM dip, enabling depth-resolved tomographic imaging \cite{Abouraddy2002}. The use of frequency-correlated photon pairs endows QOCT with several advantages over classical OCT. For instance, when the SPDC pump bandwidth is sufficiently narrow, the technique becomes inherently immune to even-order dispersion in the sample, including group-velocity dispersion (GVD) \cite{Steinberg1992, Franson1992}. Moreover, QOCT exhibits a quantum-enabled twofold enhancement in axial resolution compared with a classical OCT system of identical bandwidth \cite{Nasr2003}. In fact, axial resolutions in the sub-micrometer regime have been demonstrated through careful engineering of the biphoton source \cite{Okano2016, Katamadze2025}.

The implementation of QOCT based on analyzing the standard HOM interferogram— referred to as temporal-domain QOCT (TD-QOCT)—has generated considerable interest due to its potential applications. However, practical deployment of TD-QOCT faces several significant challenges. Chief among them are the long acquisition times needed to obtain either a full 3D tomographic data set or even a single axial (A-scan) measurement. These limitations stem from the inherently low photon flux of typical entangled-photon sources and from the need to employ mechanical stages for rasterized axial and transverse scanning.

Several works have introduced creative strategies to overcome these limitations. For instance, full-field TD-QOCT~\cite{Ibarra2020} removes the need for transverse raster scanning on the $(x,y)$ plane, substantially reducing the time required to probe a three-dimensional object. Even more significantly, by spectrally resolving the TD-QOCT interferogram of a multilayer sample---transitioning from a purely time-domain function $P(\tau)$ to a hybrid frequency-resolved function $P(\tau,\omega_s,\omega_i)$---one can extract an A-scan without mechanical scanning. This approach, known as spectral-domain QOCT (SD-QOCT), has been demonstrated in the NIR spectral region (1550 nm), showing impressive performance and reducing A-scan acquisition times by orders of magnitude compared with TD-QOCT~\cite{Yepiz-Graciano2020}. Furthermore, as will be shown here, when the SPDC source is pumped with a sufficiently narrowband laser, the SD-QOCT scheme can be simplified, enabling even faster operation and single-point A-scan acquisition on sub-second timescales.

\begin{figure}[h!]
	\centering \includegraphics[width=0.47\textwidth]{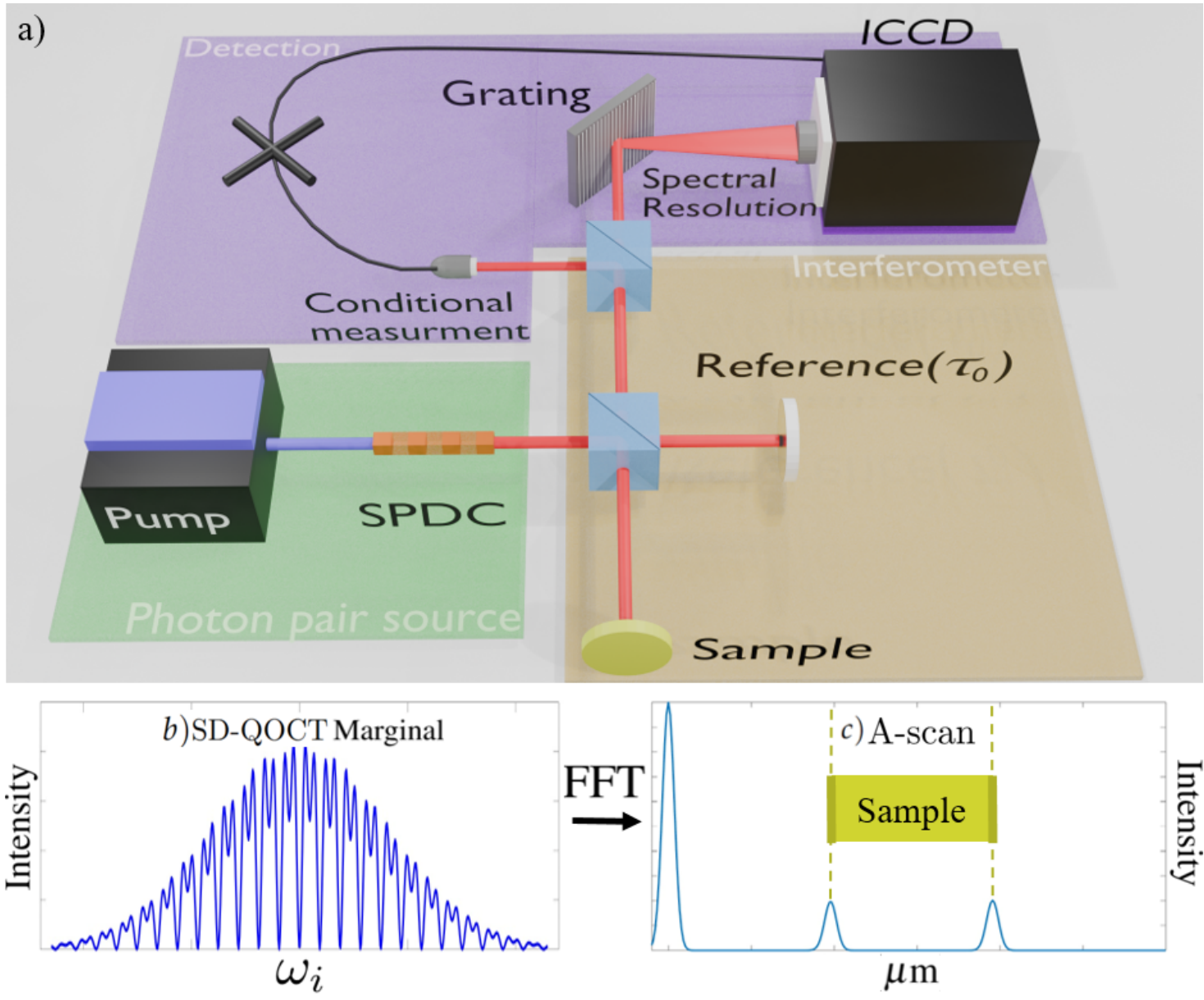}
	\caption{Schematics of the marginal SD-QOCT technique. The ICCD camera records the marginal interferogram $p_c(\tau_0,\omega_i)$  with record acquisition times, for a fixed signal--idler temporal delay $\tau_0$ selected at the reference mirror according to a specific criterion (see main text). Applying a fast Fourier transform to this marginal distribution yields the temproal-domain interferogram $p_c(\tau_0,t)$, which directly reveals the A-scan of the sample.}
	\label{fig:diagram}
\end{figure}

In this manuscript, we expand on our group’s previous work by demonstrating an optimized implementation of the SD-QOCT scheme—illustrated schematically in Fig.~\ref{fig:diagram}—capable of obtaining an A-scan in a single camera exposure with sub-second acquisition times. The method relies on extracting a one-dimensional marginal SD-QOCT interferogram, obtained by tracing over the signal-photon frequency and projecting onto a fixed temporal delay. Experimentally, this is achieved by detecting the signal photon with a bucket detector, thereby measuring the spectrum of the heralded idler photon after its interaction with the multilayer sample and subsequent Hong–Ou–Mandel interference, all at a fixed signal–idler delay.

For our proof-of-concept, we have implemented a high-flux photon-pair source based on a type-II PPKTP crystal operating in a collinear SPDC configuration in the VIS–NIR regime ($\sim 800$ nm), combined with a high-efficiency spectral analyzer comprised of a diffraction grating and an intensified CCD camera. This system enables  A-scan acquisition through a single camera exposure with a record sub-second measurement time of $100$ ms for a sample in the form of a single reflective surface (a mirror), a sub-nanometer spectral resolution of $\sim 0.05$ nm, and an operational penetration depth of $\sim 4$ mm. We believe that these results represent a significant step toward the practical deployment of optimized SD-QOCT techniques for biomedical imaging.

\section{Theory}
The two-photon state produced by the PPKTP crystal in the collinear SPDC process, when pumped with a continuous laser can be written as 

\begin{equation}
\ket{\psi} = \ket{0}_s\ket{0}_i+ \xi \int_{-\infty}^{\infty} \int_{-\infty}^{\infty} \mathrm{d}\omega_s \omega_i  f(\omega_s,\omega_i) \ket{\omega_s}\ket{\omega_i}, \label{eq:estado_mono}
\end{equation}

\noindent This state involves the energy-conserving signal $\omega_s=\omega_0+\Omega/2$ and idler $\omega_i=\omega_0-\Omega/2$ frequencies, in terms of a frequency non-degeneracy $\Omega=\omega_s-\omega_i$. Here, $\xi$ is a constant related to the conversion efficiency and $f(\omega_s,\omega_i)$ represents the joint spectral amplitude function given by 

\begin{equation}
f(\omega_s,\omega_i)= \mbox{sinc}\left( \frac{L}{2} \Delta k(\omega_s,\omega_i) \right)\mbox{exp}\left( i \frac{L}{2} \Delta k(\omega_s,\omega_i) \right),
\end{equation}

\noindent where $|f(\omega_s,\omega_i)|^2$ represents the joint spectral intensity function (JSI). 

A standard TD-QOCT scheme directs the signal and idler photons from an SPDC source to the input ports of a beamsplitter (BS) after one photon has interacted with a multilayer sample and the other has been temporally delayed, producing the characteristic sequence of Hong–Ou–Mandel (HOM) dips \cite{Hong1987} when monitoring the rate of coincidence events across the BS outputs. For an $N$-layer sample, the interferogram contains one HOM dip per physical interface, along with an additional set of $\binom{N}{2}$ artifact structures (dips or peaks) due to cross-interface interference that generally hinders accurate interpretation of the source morphology from the QOCT interferogram.

Whether an artifact associated with cross interference between two interfaces appears as a dip or a peak depends on the phase difference between the contributions from the two interfaces in question.  Recent works have proposed effective strategies to mitigate these undesired artifact structures in QOCT implementations \cite{Yepiz2023, Yepiz2025}.

A sample under study through QOCT can be represented by its reflectivity function $H(\omega)$ (SRF), which for the case of a $N$-layers sample can be written as 
\begin{equation}
H(\omega)= \sum\limits_{j=0}^{N-1} r_{j} e^{i\omega T_{j}} = r_{0}+r_{1}e^{i \omega T_{1}}+\ldots,
\end{equation}

\noindent where $r_j$ is the reflectivity of the $j$-th layer and $T_j$ is the round-trip time of flight from the first layer ($j=0$, setting $T_0=0$) to the $j$-th layer.

In TD-QOCT, the coincidence count rate at the two output ports of the BS is recorded as a function of the signal--idler delay $\tau$. This leads to the well-known HOM interferogram $P_c(\tau)$, commonly referred to as an A-scan, as it contains all relevant information about the sample's internal interfaces at a single sample transverse location. The TD-QOCT interferogram can be described by

\begin{equation}
\label{eq:Rc_tau}
\begin{split}
 P_c(\tau)= \frac{1}{2}\int_{-\infty}^{\infty} \mathrm{d}\omega_s d \omega_i | f(\omega_s,\omega_i)H(\omega_s) \\ 
-f(\omega_i,\omega_s)H(\omega_i) e^{i(\omega_s-\omega_i)\tau}|^2.
\end{split}
\end{equation}

By spectrally resolving the two photons in SD-QOCT, we retain the integrand of this last expression to yield the delay- and frequency-dependent interferogram $p_c(\tau,\omega_s,\omega_i)$, given by \cite{Yepiz-Graciano2020} 

\begin{equation}
\label{eq:rc_tau_omegas_omegai}
\begin{split}
p_c(\tau,\omega_s,\omega_i)=\frac{1}{2} |f(\omega_s,\omega_i)H(\omega_s) \\ -f(\omega_i,\omega_s)H(\omega_i) e^{i(\omega_s-\omega_i) \tau} |^2.
\end{split}
\end{equation}

From this expression, we can integrate over $\omega_s$ to obtain the two-dimensional interferogram $p_c(\tau,\omega_i)$, which can then be processed to extract information about the sample. In particular, there are two ways to obtain an A-scan from this interferogram: (i) tracing over the remaining frequency variable $\omega_i$ recovers the standard TD-QOCT interferogram $p_c(\tau)$ (the HOM dip), and (ii) fixing the temporal delay at $\tau=\tau_0$ yields the marginal interferogram $p_c(\tau_0,\omega_i)$ (see Fig.~\ref{fig:diagram}b). Applying a fast Fourier transform (FFT) to this marginal distribution produces the marginal temporal interferogram $p_c(\tau_0,t)$, which directly reveals the A-scan information (see Fig.~\ref{fig:diagram}c) \cite{Yepiz2019,KolenderskaFourierdomain2020}.

In this work, we adopt the latter approach. Our system—schematically depicted in Fig.~\ref{fig:diagram}a spectrally resolves one photon (the idler) while detecting the other one (the signal) with a bucket detector. The signal detection conditions the spectral components registered by the ICCD camera at a fixed delay. By collecting the full flux of signal photons, the bucket detector increases the effective detection rate at the ICCD, yielding higher counts per pixel in the heralded idler spectrum. This improvement allows us to obtain a complete A-scan without mechanical scanning, with integration times in the sub-second regime.


Now, in order to acquire an  A-scan without varying the temporal delay (i.e. with no mechanical scanning), we have established an empirical criterion for the fixed temporal delay $\tau_0$ at which the marginal interferogram should be recorded. It is essential to avoid delay values located within the sample's optical region, as this introduces ambiguity regarding the origin of the interference peaks—specifically, whether they stem from interfaces preceding or succeeding the reference position. This effect is evidenced in the SD-QOCT interferogram shown in Fig. 4(a). Consequently, $\tau_0$ must be selected to fulfill either $\tau_0 \ge (T_S + \sigma_{\tau})$ or the alternative condition $\tau_0 \le -\sigma_{\tau}$, where $T_S$ denotes the round-trip traversal time through the sample and $\sigma_{\tau}$ is the temporal FWHM of the HOM dip.

\section{Experiment}
\begin{figure}[h!]
	\centering \includegraphics[width=0.47\textwidth]{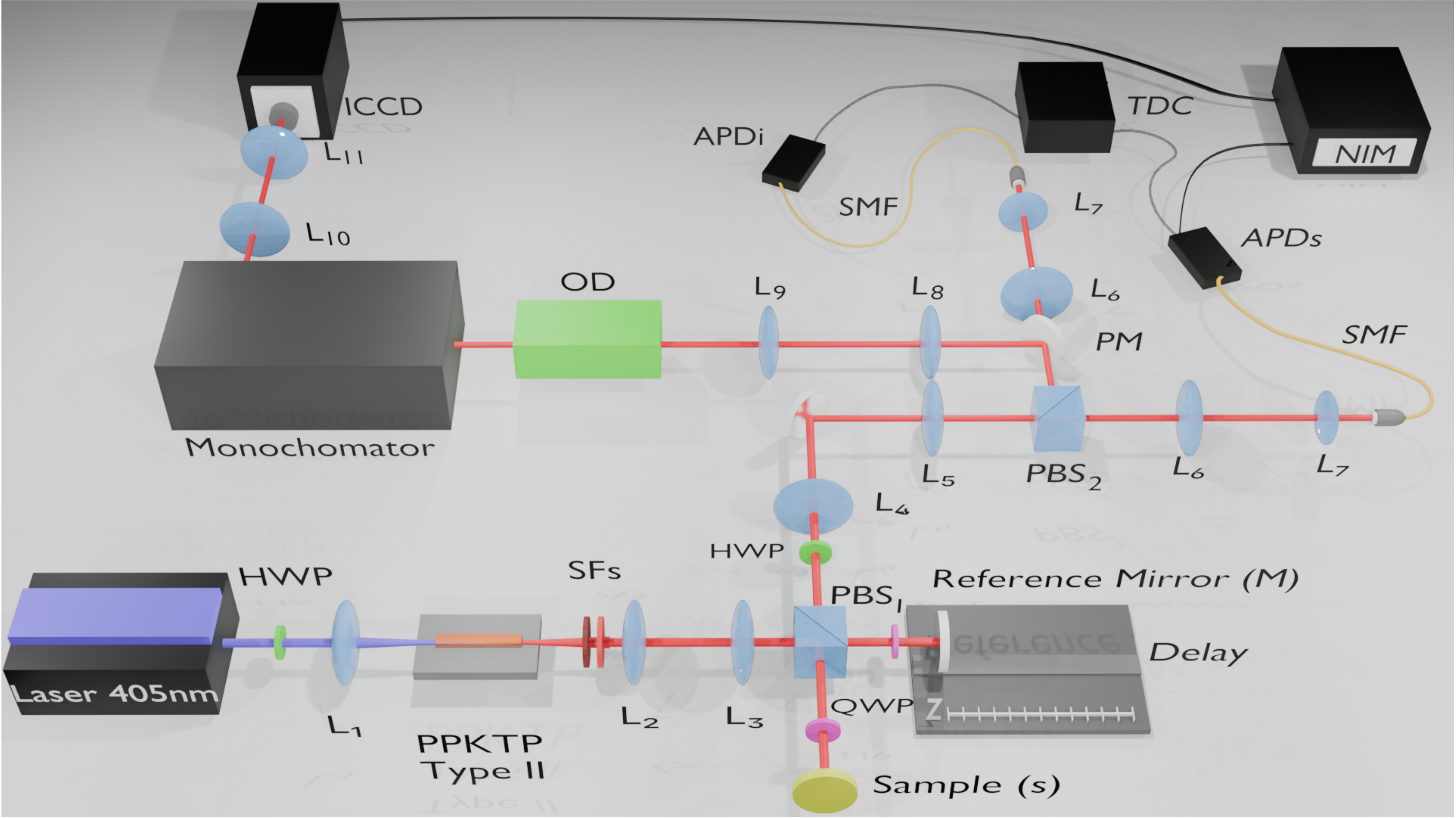}
	\caption{Experimental setup for the marginal SD-QOCT scheme, see text for description.}
	\label{fig:setup}
\end{figure}

Figure \ref{fig:setup} presents the experimental setup. A 10-mm-long type-II PPKTP crystal (10-$\mu$m poling period) is pumped with a 1-mW continuous-wave laser at 405 nm, generating cross-polarized SPDC photon pairs centered at 810 nm with a bandwidth of approximately 1 nm in a collinear configuration. The pump polarization is controlled with a half-wave plate ($HWP_1$), and the beam is focused into the crystal using lens $L_1$ ($f = 500$ mm), yielding a waist of $w_0 \approx 250~\mu$m. After the crystal, the down-converted photons propagate through a spectral filtering system (SF) consisting of a 500-nm long-pass filter and an 810 $\pm$ 10 nm band-pass filter, and are then separated by a polarizing beam splitter ($PBS_1$). The image plane of the crystal is relayed via a $1\times4f$ telescope formed by plano-convex lenses $L_2$ ($f = 100$ mm) and $L_3$ ($f = 150$ mm) onto the sample (S) and reference mirror (M) planes. The horizontally polarized photon transmitted through $PBS_1$ is directed toward the reference mirror, mounted on a translation stage with a minimum step of 10 $\mu$m. After double passage through a quarter-wave plate ($QWP_1$), its polarization is rotated to vertical, causing it to be reflected toward a second polarizing beam splitter ($PBS_2$). Conversely, the vertically polarized photon reflected at $PBS_1$ is routed to the sample plane (S)—either a mirror or a 1-mm glass slide—where it reflects, undergoes a polarization flip to horizontal after passing twice through $QWP_2$, and is subsequently transmitted through $PBS_1$ and directed to $PBS_2$.

The polarization of both photons is subsequently converted to diagonal by a half-wave plate (HWP) and relayed through a $1\times4f$ imaging system formed by plano-convex lenses $L_4$ ($f = 150$ mm) and $L_5$ ($f = 150$ mm) onto $PBS_2$, where the HOM interference takes place. The transmitted port of $PBS_2$, referred to as the signal arm, is then imaged onto the core of a single-mode fiber (SMF$_1$) using a telescope with a total demagnification of $25\times$, implemented with lenses $L_6$ ($f = 200$ mm) and $L_7$ ($f = 8$ mm). The collected light is directed to the avalanche photodiode APD$_s$. Meanwhile, the reflected port of $PBS_2$, corresponding to the idler path, is routed to either a  detection or analysis stage, as determined by a retractable mirror PM.

Reflection from PM leads to the SD-QOCT measurement stage, where the marginal interferogram $P_c(\tau_0,\omega_i)$ is obtained without the need for mechanical scanning. To achieve this, the idler photon is imaged through a $1.5\times$ telescope, formed by lenses $L_8$ ($f = 500$ mm) and $L_9$ ($f = 750$ mm), into the input port of an image-preserving optical delay line (OD) of approximately 28 m in length, providing a total propagation time of about 90 ns—sufficient to compensate for the internal delay of the ICCD camera (Andor iStar 334T) \cite{Ibarra2020}. The image plane of the idler photon exiting the OD is subsequently reduced using a $5\times$ demagnifying telescope, which forms a focused image onto the vertical entrance slit of a monochromator (Andor Shamrock 500i), where a 1200-lines/mm, 750-nm-blaze diffraction grating disperses the spectral components. At the monochromator’s output plane, and in order to enhance the overall spectral resolution, a $5\times$ magnifying telescope—constructed with lenses $L_{10}$ ($f = 50$ mm) and $L_{11}$ ($f = 250$ mm)—projects the image onto the ICCD sensor. The ICCD is triggered by the detection of the signal photon in APD$_s$; the transistor–transistor logic (TTL) pulse generated for each detection event is discriminated and appropriately delayed using a series of NIM electronic modules before reaching the camera.

The alternative path (activated when the PM is lowered) enables measurement of the standard TD-QOCT interferogram $P_c(\tau)$ (A-scan) of the object placed at the sample plane. In this configuration, the idler photon is collected into SMF$_2$ via a telescope with a total demagnification of $25\times$, implemented with the same lenses $L_6$ ($f = 200$ mm) and $L_7$ ($f = 8$ mm), and subsequently detected by APD$_i$. The A-scan is obtained by recording the coincidence rate between APD$_s$ and APD$_i$ as a function of the temporal delay $\tau$. The setup is designed to allow straightforward switching between the SD-QOCT and TD-QOCT measurement modes, for the acquisition of both types of interferogram: spectral-domain and time-domain.

\section{Results and discussion}

As an initial calibration step, we experimentally acquired the SD-QOCT interferogram $p_c(\tau,\omega_i)$ for the two samples considered in this study: a mirror and a 1~mm-thick glass slide. For the mirror sample, the interferogram was constructed from 300 individual measurements of the marginal distribution $p_c(\tau_n,\omega_i)$, each corresponding to a distinct temporal delay value $\tau_n$. For the glass slide, 500 such single-shot measurements were obtained. The delay was scanned in steps of $20~\mu\text{m}$ over a range of $6$~mm for the mirror sample and $10$~mm for the glass-slide sample. In all cases, each individual marginal measurement was recorded using the ICCD.

\begin{figure}[h!]
    \centering    \includegraphics[width=0.45\textwidth]{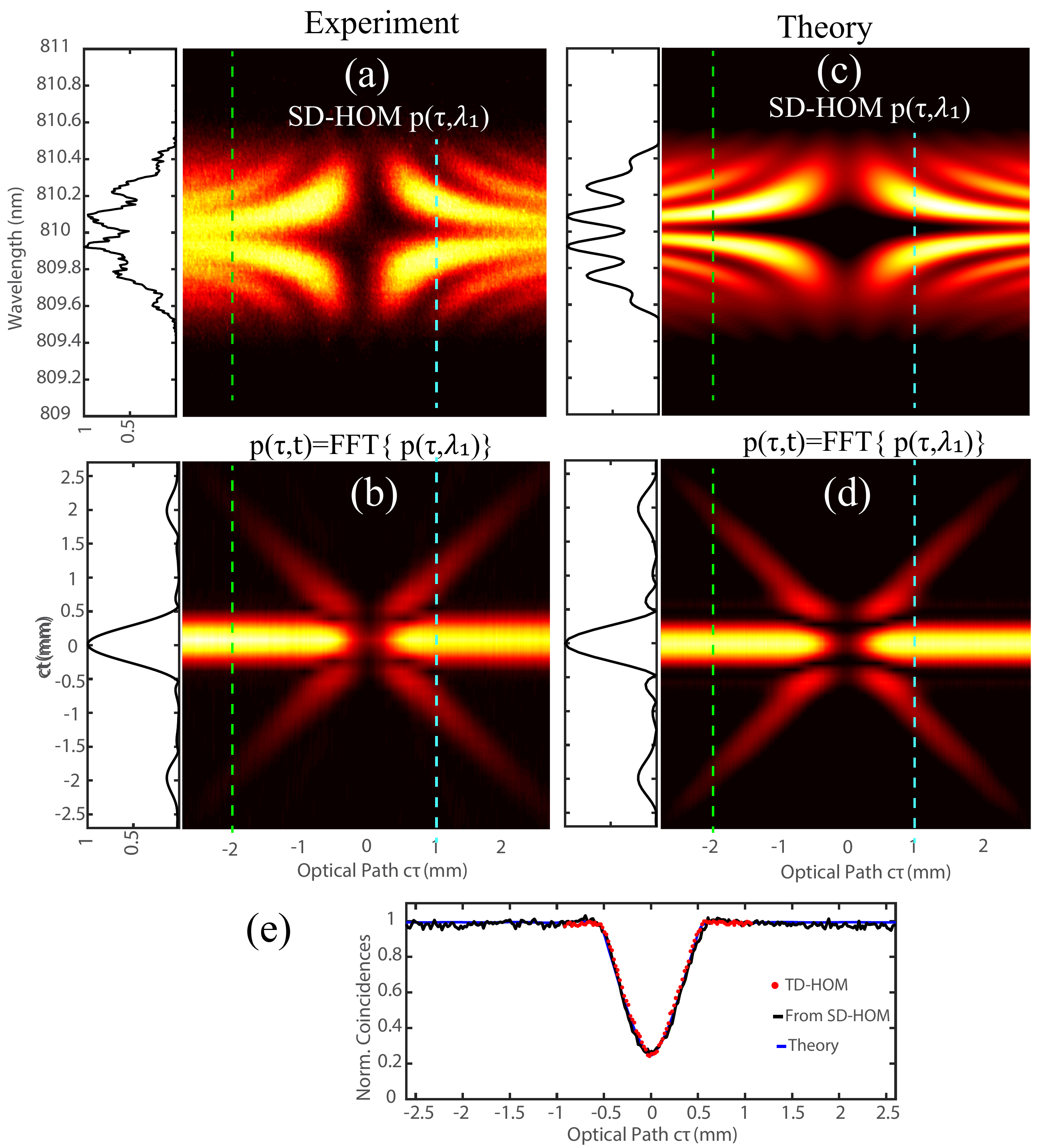}
	\caption{a) SD-QOCT interferogram for a mirror, with the corresponding time-domain interferogram (FFT) shown in panel~b. The inset on the left of panel~a illustrates a representative fixed-delay, single-exposure measurement of the marginal interferogram, acquired with a 10~s ICCD exposure at the temporal-delay value indicated by the dashed green line; its FFT is displayed in the left inset of panel~b. Panels~c and~d present the corresponding numerical simulations. Panel~e compares the TD-QOCT interferogram obtained from direct coincidence measurements (red dots), by integrating over frequency the experimental 2D interferogram in panel~a (black dots), and similarly for the numerically-obtained 2D interferogram in panel~c (blue curve).}
	\label{fig:SDQOCT-M}
\end{figure}

In Fig.~\ref{fig:SDQOCT-M} we show the full set of calibration measurements obtained for the mirror sample. The experimental SD-QOCT interferogram is presented in Fig.~\ref{fig:SDQOCT-M}a, and its corresponding temporal-domain interferogram (FFT) appears in Fig.~\ref{fig:SDQOCT-M}b. The sideplot in Fig.~\ref{fig:SDQOCT-M}a displays an example of a marginal interferogram acquired at the temporal delay value $\tau_0$, indicated by the vertical dashed green line in all panels; the FFT of this marginal distribution is shown in the sideplot of Fig.~\ref{fig:SDQOCT-M}b. The oscillatory structure of the marginal interferogram away from the center of the interferogram is clearly resolved, despite the small overall spectral range covered, which is shorter than 1 nm, a consequence of the high spectral resolution---and therefore high penetration depth---achieved by our system.

Panels c and d show the corresponding numerical simulations, including theoretical marginal interferograms evaluated at the same temporal delay values as the experimental data, demonstrating excellent agreement. Panel e displays the TD-QOCT interferogram (A-scan) obtained in three independent ways: through direct coincidence measurements (red dots), by integrating over frequency from panel a (black curve), and from panel b (blue curve). All three results show excellent agreement.

The TD-QOCT interferogram in panel e was acquired in the standard configuration by recording the coincidence rate between APD$_s$ and APD$_i$ at 100 temporal-delay positions over a 2~mm range (20~$\mu$m steps), with a 1~s integration time per delay setting.


From the measured interferogram and its corresponding FFT for the mirror sample, we can extract key performance parameters of our system, including the axial resolution and the effective imaging range. The axial resolution is given directly by the full width at half maximum (FWHM) of the HOM dip (Fig.~\ref{fig:SDQOCT-M}e), which in our setup is approximately $500~\mu\text{m}$. This relatively modest resolution---compared with previous demonstrations by our group and others \cite{Ibarra2020,Yepiz-Graciano2020,Yepiz-GracianoQuantumOptical2022,Kolenderska2025,Takeuchi2022}---stems from the narrow $\sim 1$~nm bandwidth of the photon pairs generated via collinear type-II SPDC in the PPKTP crystal. 

Although this configuration was selected for the present proof-of-concept demonstration of the marginal SD-QOCT method due to its simplicity and ease of alignment, our system can readily accommodate photon-pair sources with substantially larger bandwidths. These include type-0 PPKTP crystals ($\sim 11$~nm) or even type-I BBO crystals ($\sim 100$~nm)---the latter being the source employed in our FF-QOCT demonstration, which held the record for the highest axial resolution in QOCT until 2022 \cite{Ibarra2020,Takeuchi2022}. Future implementations of the M-SD-QOCT scheme will incorporate such broadband sources to achieve significantly enhanced axial resolution.

Additionally, the temporal-domain interferogram allows us to determine the penetration depth by applying the standard 6-dB signal-to-noise criterion commonly used in OCT. Specifically, we identify the depth at which the diagonal feature emerging from the center of the HOM dip in Fig.~\ref{fig:SDQOCT-M}b exhibits an intensity 6~dB above the noise floor. Using this criterion, our system achieves an effective penetration depth of approximately $4$~mm, representing a new record among SD-QOCT implementations (see Table~\ref{Table1}).

A final relevant parameter of our setup is the maximum measurable bandwidth, for a given fixed delay setting, which characterizes the spectral span of the idler photon that the heralded, spectrally resolved detection system---comprising the diffraction grating and ICCD---can capture in a single exposure. Our current configuration supports a usable bandwidth of $\sim 15$~nm,  with a nominal spectral resolution of $\sim 0.05$~nm.



\begin{table*}[ht]
\begin{center}
\begin{tabular}{lcccccc} 
\hline
\textbf{Technique} & \textbf{Mirror} & \textbf{Glass two-layer sample}& \textbf{Spectral res.}& \textbf{Axial res.}& \textbf{Optical depth.}  &\textbf{Ref.}\\
\hline
SD-QOCT& 10s & NA & 0.7 nm & 100 $\mu$ m & 469 $\mu$m &\cite{ZhangHighspeed2021} \\
\hline
FD-QOCT& 2s & 3min & 3.6 nm & 11 $\mu$ m & 333 $\mu$m &\cite{Kolenderska2025} \\
\hline
SD-QOCT& $\sim$ 2min & $\sim$ 2min & 1 nm & 33 $\mu$ m & 1.2 mm &\cite{Yepiz-Graciano2020} \\        
\hline
M-SD-QOCT& 0.1s & 10s & 0.05 nm  & 500 $\mu$ m & $\sim$4 mm &This work \\ 
\hline
\end{tabular}
\caption{Benchmark comparison of key parameters of SD-QOCT, for different schemes reported.}
\label{Table1}
\end{center}
\end{table*}


\begin{figure}[h!]
    \centering    \includegraphics[width=0.45\textwidth]{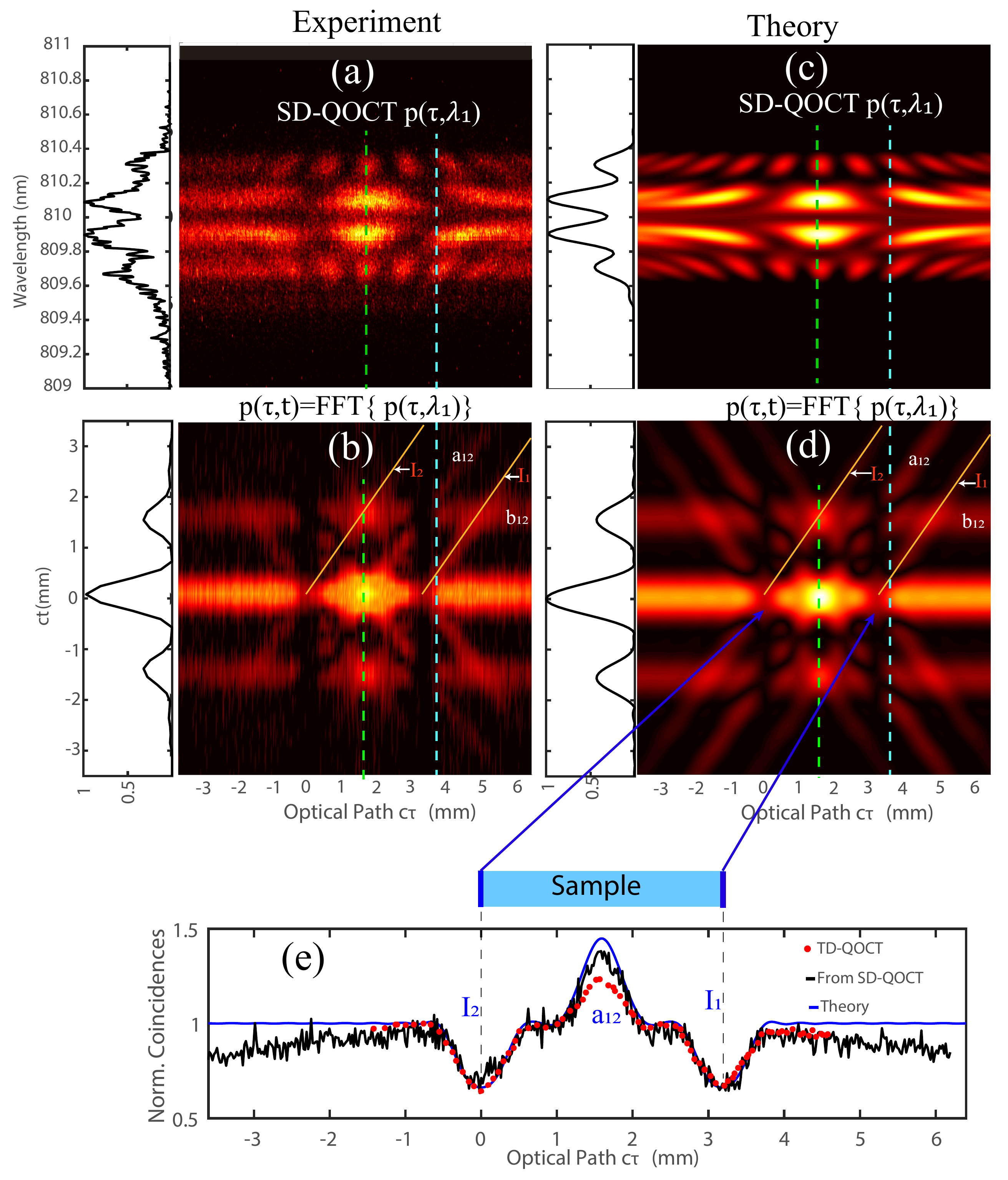}
	\caption{a) SD-QOCT interferogram for the 1~mm glass slide, with the corresponding time-domain interferogram (FFT) shown in panel~b. The sideplot in panel~a illustrates a representative fixed-delay marginal interferogram, acquired with a 10~s ICCD exposure at the temporal-delay value indicated by the dashed green line; its FFT is displayed in the sideplot of panel~b. Panels~c and~d present the corresponding numerical simulations. Panel~e compares the TD-QOCT interferogram obtained from direct coincidence measurements (red dots), by integrating over frequency the experimental interferogram in panel~a (black dots), and the theoretical interferogram in panel~c (blue curve). The colored labels in panels~b, d, and~e correspond to the distinct reflective layers and artifact features of the sample while the blue dashed line in all panels correspond to the temporal delay value used for capturing the measurements presented in Fig. \ref{fig:05MSD} (see text for details).}
	\label{fig:SDQOCT-S}
\end{figure}

In Fig.~\ref{fig:SDQOCT-S}a we show the measured SD-QOCT interferogram for the 1~mm-thick glass slide sample, with the corresponding temporal-domain interferogram displayed in Fig.~\ref{fig:SDQOCT-S}b. The sideplot in panel~a illustrates a representative marginal interferogram, acquired with a 10~s integration time at the temporal-delay value indicated by the dashed green line; its FFT is shown in the sideplot of Fig.~\ref{fig:SDQOCT-S}b. Numerical simulations are presented in panels~c and~d, together with the corresponding theoretical marginal interferograms shown in the sideplots. These simulated results exhibit excellent agreement with the experimental data.

Panel~e presents a comparison between three versions of the TD-QOCT interferogram: the directly measured coincidence data (red dots),  obtained by integrating over frequency in the experimental data (black curve), and similarly for the numerical simulation (blue curve). All three traces exhibit very good agreement. In this measurement, both physical interfaces of the glass slide are clearly resolved, together with the central artifact peak characteristic of multilayer samples. The slight offset observed in the baseline coincidence level between experiment and theory arises from small misalignments at large temporal delays, i.e., far from the center of the HOM dip.

The colored labels in panels~b, d, and~e identify the main structural features of the interferogram: the first interface (I$1$), the second interface (I$2$), and the characteristic artifact features (a${12}$ and b${12}$). These structures together provide complete information about the interfaces and reflective layers of the sample in a single measurement. The TD-QOCT interferogram in panel~e was obtained in the standard configuration by recording coincidences between APD$_s$ and APD$_i$ over 300 temporal-delay positions across a 6~mm range (20~$\mu$m steps), with a 1~s integration time per point.

\begin{figure}[h!]
    \centering    \includegraphics[width=0.47\textwidth]{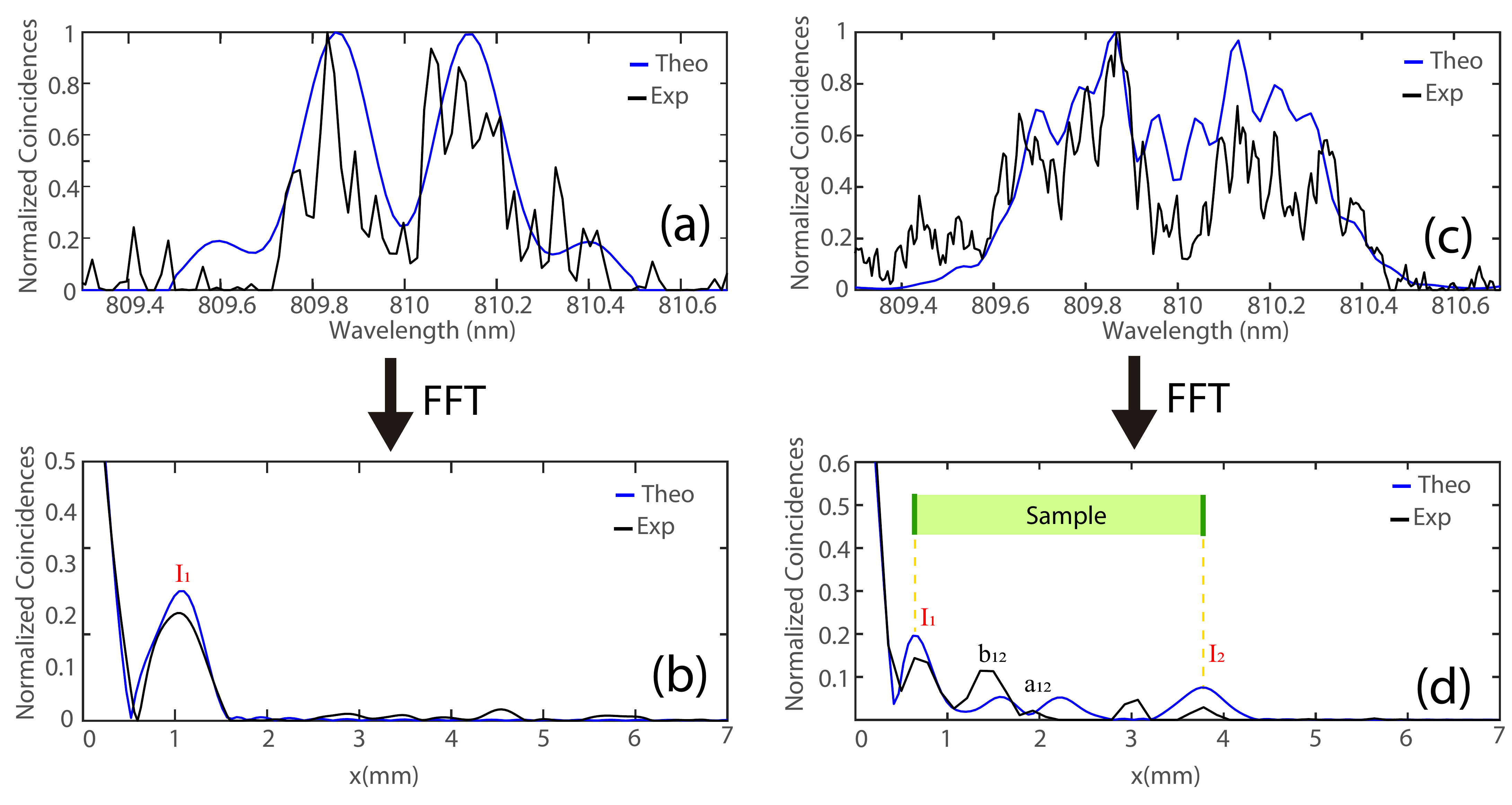}
	\caption{Measurements of the marginal SD-QOCT interferogram $P_c(\tau_0,\omega_i)$ obtained at a fixed delay value in a single camera exposure: a) mirror sample, acquired with a 0.1~s ICCD integration time, and b) 1~mm glass slide, acquired with a 10~s integration time. The lower panels show the corresponding FFTs, revealing the reflective surface of the mirror in panel~c and the two interfaces of the glass slide in panel~d, together with a schematic representation of the sample. The blue curves in all four panels show the numerical simulations, which exhibit excellent agreement with the experimental data.}
	\label{fig:05MSD}
\end{figure}

Finally, Fig.~\ref{fig:05MSD}a and Fig.~\ref{fig:05MSD}b show the recorded marginal interferograms for the mirror and the glass-slide samples, acquired at temporal-delay positions of $c\tau_0 = 1$~mm (blue dashed line in Fig. \ref{fig:SDQOCT-M}) for the mirror and $c\tau_0 = 3.8$~mm for the glass slide (blue dashed line in Fig. \ref{fig:SDQOCT-S}). These delay values satisfy the fixed-delay single exposure measurement  criterion $c\tau_0 \ge cT_s + c\sigma_\tau$, where in our case $c\tau_0 = 3.8$~mm, $cT_s = 3.2$~mm, and $c\sigma_\tau = 0.5$~mm, all within the measured penetration depth of approximately 4~mm. The FFTs of these marginal interferograms are shown in Fig.~\ref{fig:05MSD}c and Fig.~\ref{fig:05MSD}d, corresponding to the reconstructed A-scans. The reflective surface of the mirror, as well as the front and rear interfaces of the glass slide, are clearly resolved at the expected optical-path differences based on the sample thickness. The blue curves in each panel show the corresponding numerical simulations, which agree exceptionally well with the experimental data.

The marginal interferograms shown were obtained in a single ICCD exposure, with acquisition times of 100ms for the mirror and 10~s for the glass slide, establishing new record acquisition times for A-scan measurements in SD-QOCT, as summarized in Table~\ref{Table1}.

Importantly, as already mentioned, our system allows for a straightforward and significant upgrade: the axial resolution could be enhanced almost seamlessly to $\sim 11~\mu\text{m}$ by replacing the SPDC source with a type-0 PPKTP crystal \cite{Kolenderska2025}. Additionally, although implementing artifact-removal strategies is beyond the scope of the present work, several of the techniques reported in the literature are fully compatible with our approach. These include phase-compensation methods that suppress artifacts via destructive interference, as well as optimization strategies based on genetic algorithms to identify and eliminate unwanted features in the interferogram \cite{Yepiz2019,Yepiz2025}. We plan to incorporate such techniques in future optimized implementations of the M-SD-QOCT scheme.

\section{Conclusion}
To summarize, we have demonstrated the marginal spectral-domain quantum optical coherence tomography (M-SD-QOCT) scheme, which extracts a one-dimensional marginal measurement from the full three-dimensional SD-QOCT interferogram. This strategy greatly simplifies the experimental implementation and enables fixed-delay, single-exposure acquisition of an A-scan with sub-second integration times. Our setup employs a diffraction grating, a high-resolution intensified CCD camera, and a high-flux photon-pair source operating in the VIS--NIR range ($\sim 800$~nm). The technique yields a single transverse A-scan with record acquisition times of 0.1~s for a mirror sample and 10~s for a 1~mm glass-slide sample, with a sub-nanometer spectral resolution of $\sim 0.05$~nm and an effective penetration depth of approximately 4~mm. Importantly, the measured results are in excellent agreement with the corresponding theoretical predictions. We believe these capabilities represent a significant step toward establishing SD-QOCT as a practical tool for biomedical imaging.


{\bf Funding.} This work was supported by SECIHTI, México (CBF-2025-I-2699, BP-BSNAC-20250429132744996-10682132).

{\bf Acknowledgments} 
D. S. R. (CVU 1345671) and P. Y. G. (CVU 449757) acknowledge financial support from SECIHTI (formerly CONAHCYT) Mexico.

{\bf Disclosures} 
The authors declare no conflicts of interest.

{\bf Data Availability Statement} 
Data underlying the results presented in this paper are not publicly available at this time but may be obtained from the authors upon reasonable request.

\bibliographystyle{apsrev4-2}
\bibliography{references}

\end{document}